\documentclass[prb,showpacs,preprintnumbers,amsmath,amssymb]{revtex4}
\usepackage{bm}
\usepackage[dvips]{graphicx}
\begin{document}

\title{ Long Range Dynamics Related to Magnetic Impurity in the 2D Heisenberg Antiferromagnet}
\author{O.P. Sushkov}
\email{sushkov@newt.phys.unsw.edu.au}

\affiliation{Max-Planck-Institute f\"ur Festk\"orperforschung,
Heisenbergstrasse 1, D-70569 Stuttgart, Germany}
\affiliation{School of Physics, University of New South Wales, Sydney 2052,
Australia.}


\begin{abstract}
We consider a magnetic impurity in the two-dimensional Heisenberg
antifferomagnet with long range antiferromagnetic order. At low temperature
the impurity magnetic susceptibility
has a Curie term ($\propto 1/T$) and a logarithmic correction ($\propto \ln(T)$).
We calculate the correction and derive related Ward identity for the
impurity-spin-wave vertex.
\end{abstract}
 \maketitle

\section{introduction}
The problem of magnetic impurities interacting with a system of strongly
correlated electrons has attracted much interest recently, mainly due to the
experimental discoveries of the high-T$_c$ superconductors and new heavy
fermion compounds. In the field of the high-T$_c$ materials, the parent
compounds are known to be two-dimensional (2D) antiferromagnetic (AFM)
Mott-Hubbard insulators based on CuO$_2$ planes
which are driven to a superconducting state by doping
(e.g. with holes) \cite{Man,Dag}.
Even though the holes can hop, thus destroying the AFM long-range order and
causing the development of superconducting pairing, the extreme limit of
static holes also has a physical relevance.
Systems with static holes have been also realized experimentally
 in cuprates \cite{Cheong,Vajk}

Several theoretical studies have addressed isolated static holes 
\cite{Nag,Bulut,Sand1,Sachd,Sand2} and added spins \cite{Igar,Kotov,Sand2} in
2D Heisenberg antiferromagnets with long range AFM order.
A singular logarithmic frequency behavior of the perpendicular magnetic susceptibility 
at zero temperature has been found in Ref.\cite{Nag}, see also a discussion in Ref.\cite{Chern}.
A very interesting problem is a magnetic impurity in 2D Heisenberg
antiferromagnet at O(3) quantum critical point \cite{Sachd,Sus,Troy}.
However, this problem is out of the scope of the present paper.
The low temperature behavior of the impurity static magnetic susceptibility in a
gapped system is trivial, it obeys the simple Curie low, $\delta\chi
=\Omega(\Omega+1)/3T$, where $\Omega$ is the impurity spin.
However, for 2D systems which possess the long range AFM order at zero temperature,
the excitation spectrum  is gapless due to Goldstone spin waves and the Curie low is not obvious.
A very interesting prediction \cite{Sachd} for such a regime is the classical Curie
low, $\delta\chi=\Omega^2/3T$. The behaviour is classical because of alignment of the impurity moment
with the local Neel order. This behaviour has been recently confirmed in Monte Carlo
simulations for 2D $S=1/2$ Heisenberg antiferromagnetic clusters with magnetic
impurity  \cite{Sand2}.
Moreover, in these simulations a logarithmic correction, $\propto \ln(T)$, to
the classical Curie low has been found. Both the classical Curie low
and the logarithmic
correction are related to the nontrivial long-range dynamics in the system.
In the present work we calculate the logarithmic correction using two
different methods, a) Spin-wave perturbation theory, b) Semiclassical
non-linear $\sigma$-model. In the leading $1/S$ approximation both methods
give the same result. However the results must be identical in all orders in
$1/S$ and hence the comparison allows us to derive the Ward identity for the
impurity-spin-wave vertex. Value of the logarithmic correction to the magnetic
susceptibility is in agreement with Ref. \cite{Sand2}.

A crossover from quantum to classical Curie low for a
finite AFM cluster with impurity is discussed in Section II. 
In Section III we derive the impurity susceptibility using the spin-wave
perturbation theory, and in Section IV we obtain the same result using the
semiclassical non-linear $\sigma$-model and derive the Ward identity.

\section{Curie term}
The Hamiltonian of the system under consideration is
\begin{eqnarray}
\label{H}
&&H=H_0+H_{int}+H_B\ , \nonumber\\
&&H_0=J\sum_{<ij>}{\bf S}_i\cdot{\bf S}_j\ ,\nonumber\\
&&H_{int}= J_{\perp} {\bm S}_0\cdot{\bm \Omega} \ , \nonumber\\
&&H_B= -{\bf B}\cdot\left({\bm \Omega}+\sum_i{\bm S}_i\right)
\end{eqnarray}
where ${\bm S}_i$ is spin 1/2 at the site $i$ of square lattice
with antiferromagnetic interaction ($J > 0$), 
${\bm \Omega}$ is the impurity spin coupled to the  lattice spin at
site $0$, and ${\bm B}$ is magnetic field. To be specific we will assume that 
$J_{\perp} > 0$, but all 
results are in the end independent of sign of $J_{\perp}$.
Consider an  $L\times L$ cluster ($L \gg 1 $ is even)
described by the Hamiltonian $H_0$, so no impurities for the beginning.
Energy spectrum of the system is known very well \cite{Neub,Runge,Has}.
Spin of the ground state is 0, and lowest excitations are described by
rotational spectrum of the solid top (diatomic molecule with zero projection
of spin on axis of the molecule. $K=0$)
\begin{equation}
\label{S0}
E_{\cal J}=\frac{{\cal J}({\cal J}+1)}{2I},
\end{equation}
where ${\cal J}=0,1,2...$ is spin of the state (${\cal J}=0$ corresponds to the ground
state),  $I=L^2\chi_{\perp}$ is moment of inertia of the top, and 
$\chi_{\perp}\approx 0.066/J$ is perpendicular magnetic susceptibility
\cite{Singh,Zheng}.
The spectrum (\ref{S0}) is valid as soon as the rotation is solid, i.e. 
internal degrees of freedom of the top are not excited. The first internal
excitation is the spin wave with wave length $\lambda=L$ (periodic boundary
condition). Energy of this excitation is
\begin{equation}
\label{esw}
\Delta_{sw}= 2\pi c/L, \ \ \ c \approx \sqrt{2}J.
\end{equation}
There are 8 degenerate spin-wave excitations: $S_z=\pm 1$, $x$ and $y$
directions,  and two excitations ($\cos$ and $\sin$) in each direction.
If we consider $T\ll \Delta_{sw}$ then only rotations (\ref{S0}) are important.
A more accurate criterion for solid rotation is: $8\exp(-\Delta_{sw}/T) \ll 1$, i.e.
\begin{equation}
\label{Tsw}
T \ll T_{sw} \approx \frac{\Delta_{sw}}{\ln(8)}.
\end{equation}
In this temperature regime magnetic susceptibility of the cluster is
determined by the spectrum (\ref{S0}).
\begin{eqnarray}
\label{ms0}
\chi_0&=&\frac{\partial}{\partial B}
\frac{\sum_{{\cal J,J}_z} {\cal J}_ze^{-\frac{E_{\cal J}-{\cal J}_zB}{T}}}
{\sum_{{\cal J,J}_z} e^{-\frac{E_{\cal J}-{\cal J}_zB}{T}}}=
\frac{1}{3T}\left(
\frac{\sum_{\cal J} ({\cal J}+1/2)^3 e^{-\frac{({\cal J}+1/2)^2}{2IT}}}
{\sum_{\cal J}({\cal J}+1/2) e^{-\frac{({\cal J}+1/2)}{2IT}}}
-\frac{1}{4}\right) \ ,
\end{eqnarray}
where summation over ${\cal J}$ is performed from $0$ to $\infty$.
If $ \ T \ll T_{rot}=1/I$ the susceptibility is zero, and if $T_{rot} \ll T \ll
T_{sw}$ the evaluation of (\ref{ms0}) gives
\begin{equation}
\label{c0}
\chi_0=\frac{1}{3T}\left(2IT-\frac{1}{3}+\frac{1}{18IT}+...\right)=\frac{2}{3}L^2\chi_{\perp}-
\frac{1}{9T}\left(1-\frac{T_{rot}}{6T}+...\right).
\end{equation}
The expansion is in integer powers of $T_{rot}/T$.

Now let us put an impurity with spin ${\bm \Omega}$ which interacts with
cluster via $H_{int}$, see Eq. (\ref{H}).
Excitation spectrum of the cluster with impurity is slightly different from (\ref{S0}).
Now this is a symmetric top with spin projection on axis of the top $K=\Omega$.
This is like a diatomic molecule with uncompensated electron spin and strong spin-axis
interaction. Rotational spectrum of such top is \cite{Land}
\begin{equation}
\label{J}
E_{\cal J}=\frac{{\cal J}({\cal J}+1)-2\Omega^2+<{\bm \Omega}^2>}{2I},
\end{equation}
where ${\cal J}=\Omega, \Omega+1,\Omega+2,...$ is total spin of the cluster,
and $<{\bm \Omega}^2>$ is the average value of ${\bm \Omega}^2$ in the
intrinsic reference frame. 
Similar to the previous case the spin waves are not excited as soon as the
inequality (\ref{Tsw}) is valid.
One can easily check that in this case  the magnetic susceptibility of
the cluster $\chi_1$ is given by the same Eq. (\ref{ms0}), the only difference is that
summation over ${\cal J}$ is performed not from $0$ to $\infty$, but
from $\Omega$ to $\infty$.
If $T \ll T_{rot}$ the cluster susceptibility is $\chi_1=\Omega(\Omega+1)/3T$.
If $T_{rot} \ll T \ll T_{sw}$ evaluation of (\ref{ms0}) gives
\begin{equation}
\label{c1}
\chi_1=\frac{1}{3T}\left(2IT-\frac{1}{3}+\Omega^2+\frac{1+3\Omega^2}{18IT}+...\right)=
\frac{2}{3}L^2\chi_{\perp}-\frac{1-3\Omega^2}{9T}+\frac{1+3\Omega^2}{18IT^2}+...
\end{equation}
Hence the impurity susceptibility defined as $\chi_1-\chi_0$ reads
\begin{equation}
\label{imp}
\chi_{imp}=\chi_1-\chi_0=\frac{\Omega^2}{3T}\left(1+\frac{T_{rot}}{2T}+...\right)
\end{equation}
The expansion goes in integer powers of $T_{rot}/T$. The leading term in (\ref{imp})
agrees with \cite{Sachd,Sand2}.

It is interesting to note that if there are n impurities with spin $\Omega$ on the same
sublattice than in Eq. (\ref{imp}) one shall replace
$\Omega\to n\Omega$. The impurities are not independent because the
cluster is rigid. In the thermodynamic limit it means that all the impurities
within the correlation length are not independent.

\section{logarithmic correction to susceptibility, the spin-wave derivation}

The above consideration is valid for the case when the inequality (\ref{Tsw})
is valid and hence there are no {\it real} spin-wave excitations. However
virtual spin-wave excitations  are always there. In the intrinsic reference
frame the cluster looks like picture in Fig.1

\begin{figure}[h]
\vspace{0pt}
\centering
\includegraphics[height=40pt,keepaspectratio=true]{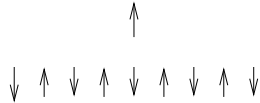}
\caption{\it Schematic picture of an AFM cluster with impurity}
\label{Fig1}
\end{figure}
\noindent
The interaction Hamiltonian $H_{int}$, Eq. (\ref{H}), can be rewritten in
terms of spin-wave 
operators $a$ and $b$ (for review of the spin-wave theory see, e. g. Ref. \cite{Man})
\begin{eqnarray}
\label{hint1}
H_{int}&=&J_{\perp}{\bm \Omega}\cdot{\bm S}_0\to 
\frac{1}{2}J_{\perp}\left(\Omega_+S_{0-}+\Omega_{-}S_{0+}\right) =
\frac{1}{2}J_{\perp}\left(\Omega_+b_0+\Omega_{-}b_0^{\dag}\right)\nonumber\\
&=&\frac{1}{2}J_{\perp}\sqrt{\frac{2}{L^2}}\sum_q\left(\Omega_+b_q+\Omega_{-}b_q^{\dag}\right) =
\frac{1}{\sqrt{2L^2}}J_{\perp}\sum_q\left(\Omega_+[u_q\beta_q+v_q\alpha_{-q}^{\dag}]+
\Omega_{-}[u_q\beta_q^{\dag}+v_q\alpha_{-q}]\right)\nonumber\\
&\to&
\frac{1}{\sqrt{2L^2}}J_{\perp}\sum_qZ_{\Gamma}\left(\Omega_+[u_q\beta_q+v_q\alpha_{-q}^{\dag}]+
\Omega_{-}[u_q\beta_q^{\dag}+v_q\alpha_{-q}]\right).
\end{eqnarray}
Here $\beta_q^{\dag}$ and $\alpha_q^{\dag}$ are creation operators for spin waves
with spin projection $S_z=\pm 1$ respectively, $u_q$ and $v_q$ are Bogoliubov parameters,
\begin{eqnarray}
\label{uv}
u_q^2=\frac{J}{\omega_q^{(0)}}+\frac{1}{2} \ ,\ \ \ 
v_q^2=\frac{J}{\omega_q^{(0)}}-\frac{1}{2} \ ,
\end{eqnarray}
and $\omega_q^{(0)}=2J\sqrt{1-\gamma_q^2} \to \sqrt{2}J q$ is the spin wave
dispersion in the leading $1/S$ approximation. In (\ref{hint1}) we assume that $S=1/2$.
In Eq. (\ref{hint1}) the impurity-spin-wave vertex is derived in the leading
$1/S$ approximation. However, at the last step we have introduced the vertex
renormalization factor $Z_{\Gamma}$ that takes into account all higher $1/S$
corrections to the vertex. Generally speaking $Z_{\Gamma}$  depends on $q$,
but here we consider only the small $q$ limit, $Z_{\Gamma}=Z_{\Gamma}(q=0)$.

To use perturbation theory we will assume that $J_{\perp} \ll J$. 
One loop correction to the impurity energy is given by  diagram in Fig.2.
\begin{figure}[h]
\vspace{0pt}
\centering
\includegraphics[height=40pt,keepaspectratio=true]{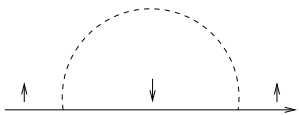}
\vspace{0pt}
\caption{\it {One loop correction to the impurity energy, the solid line shows
the impurity and the dashed line shows virtual spin wave.}}
\label{Fig2}
\end{figure}
\noindent
The diagram is infrared convergent because it contains the denominator,
$\Delta\epsilon =\epsilon_{\uparrow}-\epsilon_{\downarrow}=J_{\perp}<S_z>$, related to
the flip of the impurity spin. Here $<S_z>=\frac{1}{2}Z_s\approx 0.307$
is staggered magnetization of the lattice, and
$Z_s \approx 0.61$ is the renormalization factor for the staggered
magnetization \cite{Man,Singh,Zheng}.

Interaction of the impurity with perpendicular magnetic field is of the form 
\begin {equation}
\label{hb}
H_B=-B\Omega_x=-\frac{1}{2}B(\Omega_++\Omega_-) \ .
\end{equation}
The leading contribution to the impurity energy related to $B$ is given by diagram in Fig.3.
\begin{figure}[h]
\vspace{0pt}
\centering
\includegraphics[height=20pt,keepaspectratio=true]{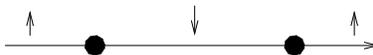}
\vspace{0pt}
\caption{\it {Leading contribution to the impurity energy related to magnetic field.
The dot denotes interaction with magnetic field.}}
\label{Fig3}
\end{figure}
\noindent
Corresponding formula reads
\begin{equation}
\label{loop0}
\delta\epsilon_0=-\frac{(B/2)^2}{\Delta\epsilon} \ .
\end{equation}
The contribution is finite because $\Delta\epsilon$ is finite. Now let us look at one loop
corrections to $\delta\epsilon_0$ shown in Fig.4.
\begin{figure}[h]
\vspace{0pt}
\centering
\includegraphics[height=55pt,keepaspectratio=true]{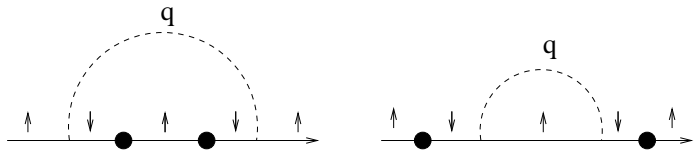}
\vspace{0pt}
\caption{\it One loop spin-wave corrections to the impurity energy related to
the magnetic field. The dot denotes interaction with the field.}
\label{Fig4}
\end{figure}
\noindent
These corrections are infrared divergent because they contain the energy
denominator without spin flip, $\epsilon_{\uparrow}-\epsilon_{\uparrow}$.
Using usual Schroedinger perturbation theory one finds the
corresponding energy correction
\begin{equation}
\label{loop1}
\delta\epsilon_1=-{(\sqrt{2\Omega}B/2)^2}\sum_q
\left(\frac{\left(\frac{\sqrt{2\Omega}J_{\perp}Z_{\Gamma}}
{\sqrt{2L^2}}u_q\right)^2}{(\Delta\epsilon+\omega_q)^2\omega_q}
+\frac{\left(\frac{\sqrt{2\Omega}J_{\perp}Z_{\Gamma}}
{\sqrt{2L^2}}v_q\right)^2}{(\Delta\epsilon)^2\omega_q}\right).
\end{equation}
Here $\omega_q\to cq =Z_c\sqrt{2}J q$, where $Z_c \approx 1.17$ is the
spin-wave velocity renormalization due to higher $1/S$ corrections
\cite{Man,Singh,Zheng}. 
Keeping only divergent terms we find from (\ref{loop1}) the following expression
for the susceptibility
\begin{equation}
\label{loop1c}
\delta\chi_{\perp}\to
\frac{J_{\perp}^2Z_{\Gamma}^2\Omega^2}{L^2(\Delta\epsilon)^2}
\sum_q\frac{u_q^2+v_q^2}{\omega_q}=
\frac{J_{\perp}^2Z_{\Gamma}^2\Omega^2}{Z_c(\Delta\epsilon)^2}
\int\frac{1}{q^2}\frac{d^2 q}{(2\pi)^2}
=\frac{Z_{\Gamma}^2\Omega^2}{2\pi Z_c J<S_z>^2}\int\frac{dq}{q}
\end{equation}
It is interesting that (\ref{loop1}) is independent of $J_{\perp}$.
We have to put some lower limit in the integral in (\ref{loop1c}).
At low temperature, when inequality (\ref{Tsw}) is valid, we have to
substitute the spin-wave gap (\ref{esw}) as the lower limit.
This gives temperature independent contribution to the impurity susceptibility
\begin{equation}
\label{l1c}
\delta\chi_{\perp}=\frac{Z_{\Gamma}^2\Omega^2}{2\pi Z_cJ <S_z>^2}\ln(L)
=\frac{2\Omega^2Z_{\Gamma}^2}{\pi Z_c Z_s^2}\ln(L)
\end{equation}
If $J\gg T \gg T_{sw}$ one has to put temperature as the lower
limit in (\ref{loop1c}). Hence
\begin{equation}
\label{l1ct}
\delta\chi_{\perp}=\frac{2\Omega^2Z_{\Gamma}^2}{\pi Z_c Z_s ^2 J}\ln(J/T)
\end{equation}
In the leading $1/S$ approximation one has to set $Z_{\Gamma}=Z_c=Z_s=1$,
hence $\delta\chi_{\perp} \to \frac{2\Omega^2}{\pi J}\ln (..)$.
This corresponds to the low-frequency susceptibility derived in Ref. \cite{Nag}.
We have calculated the perpendicular susceptibility (\ref{l1c}), (\ref{l1ct})
in the intrinsic reference frame. The isotropic susceptibility is related to
$\delta\chi_{\perp}$ by the standard relation, $\delta\chi=\frac{2}{3}\delta\chi_{\perp}$.

\section{Non-linear $\sigma$-model derivation of the log correction, and
Ward identity for the impurity spin-wave vertex}

An alternative derivation of (\ref{l1c}) and (\ref{l1ct}) is based on the $\sigma$-model.
Let us consider a field ${\bf n}$, $|{\bf n}|=1$, defined on  a disc of radius $L$.
An impurity with spin $\Omega$ is in the center of the disc.
\begin{figure}[h]
\vspace{0pt}
\centering
\includegraphics[height=130pt,keepaspectratio=true]{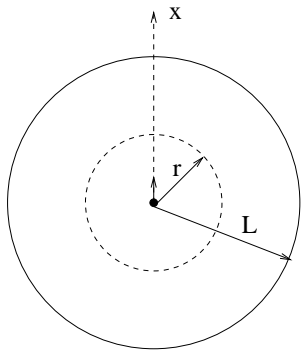}
\vspace{0pt}
\caption{\it {An impurity (dot) in the center of a disc of radius $L$. The
    disc carries fild ${\bm n}$ described by the  nonlinear
$\sigma$-model. Due to magnetic field $B=B_x$ the impurity spin is tilted in x-direction
}}
\label{Fig5}
\end{figure}
\noindent
The impurity spin is directed along z axis (perpendicular to the plane) and due
to the magnetic field $B=B_x$ it is tilted by angle $\theta$ is x-direction,
see Fig. 5. Energy of the medium is
\begin{equation}
\label{sigma}
E_{\sigma}=\frac{1}{2}\rho_s\int\left(\nabla {\bf n}\right)^2 d^2r \ ,
\end{equation}
where $\rho_s= Z_{\rho}\frac{J}{4}$ is the spin stiffness. Here $\frac{J}{4}$ is
the leading $1/S$ value for the stiffness, and $Z_{\rho}\approx 0.72$ is the
renormalization factor due to higher $1/S$-corrections \cite{Man,Singh,Zheng}.
The field is of the form ${\bf n}\approx (n_x,0,1)$. Due to
(\ref{sigma}) the field $n_x$ obeys the usual Poisson equation,  solution of
the equation is
\begin{equation}
\label{lapl}
n_x(r)=a\ln(L/r) \ .
\end{equation}
To find the constant $a$ we have to remember that $n_x(r\sim 1) =\theta$, where
$\theta$ is tilting angle of the impurity, and $1=lattice \ spacing$. 
Therefore $a=\theta/\ln(L)$.
Substituting (\ref{lapl}) in (\ref{sigma}) we find the elastic energy
$E_{\sigma}=\pi\rho_s\theta^2/\ln(L)$.
Total energy related to the impurity consists of the magnetic energy and the elastic
energy
\begin{equation}
\label{s1}
E=-B\ \Omega \ \theta+\frac{\pi\rho_s\theta^2}{\ln(L)} \ .
\end{equation}
Minimizing it with respect to $\theta$ we find
$\theta=B\Omega\ln(L)/(2\pi\rho_s)$, magnetic moment $M=\Omega\theta$, and the
magnetic susceptibility
\begin{equation}
\label{s2}
\delta\chi_{\perp}=\frac{\Omega^2}{2\pi \rho_s}\ln(L)\to \frac{2\Omega^2}{\pi
  Z_{\rho}J}\ln(L) \ .
\end{equation}
At the final step we have substituted  $\rho_s=J Z_{\rho}/4$.
In the leading $1/S$ approximation, $Z_{\Gamma}=Z_c=Z_s=Z_{\rho}=1$, hence
equation (\ref{s2}) agrees with the spin-wave results (\ref{l1c}) and (\ref{l1ct}).
Moreover a comparison of these equations gives a nontrivial Ward identity
relating renormalization factors for the spin-wave vertex $Z_{\Gamma}$,
the spin-wave velocity $Z_c$, the staggered magnetization $Z_s$, and
the spin stiffness $Z_{\rho}$
\begin{equation}
\label{w}
\frac{Z_{\Gamma}^2}{Z_cZ_s^2}=\frac{1}{Z_{\rho}} \ .
\end{equation}
This gives previously unknown value of $Z_{\Gamma}$
\begin{equation}
\label{zg}
Z_{\Gamma}=\sqrt{\frac{Z_c Z_s^2}{Z_\rho}} \approx 0.76 \ . 
\end{equation}
Similar to (\ref{l1ct}) one has to substitute $\ln(J/T)$ instead of $\ln(L)$
in (\ref{s2}) if $J \gg T > T_{sw}$. If the external magnetic field has a
nonzero frequency $\omega$, and $\omega > T, T_{sw}$, then $\ln(L) \to \ln(J/\omega)$.
Equations (\ref{l1c}), (\ref{l1ct}), and (\ref{s2}) are in agreement with
Ref. \cite{Nag} and with   recent results \cite{Sand2,Sachd1}.

The spin-wave approach in section III assumes that $J_{\perp} \ll J$. 
There are numerous two-loop diagrams which are proportional to
$(J_{\perp}/(2\pi)^2J^2)\ln(L)$ and even to $(J_{\perp}/(2\pi)^2J^2)\ln^2(L)$.
Some of the diagrams which contain the logarithm squared are shown in Fig. 6.
Calculation of all the diagrams is a pretty involved problem.
\begin{figure}[h]
\vspace{0pt}
\centering
\includegraphics[height=60pt,keepaspectratio=true]{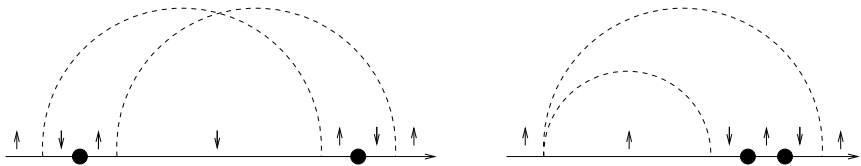}
\vspace{0pt}
\caption{\it {Two loop double logarithmic contributions to
    $\delta\chi_{\perp}$  proportional to $J_{\perp}/J^2.$
}}
\label{Fig6}
\end{figure}
\noindent
On the other hand the semiclassical derivation based on the $\sigma$-model
is independent of $J_{\perp}/J$. The only assumption in the derivation is that
the impurity magnetic moment is localized in the vicinity of the impurity.
To check this assumption we have calculated  the magnetic cloud around the
impurity using the spin-wave theory. We have found that density of the induced
magnetization $\delta\mu_z$ drops down faster than $1/r^2$, $\delta\mu_z(r) <
1/r^2$. Therefore, the magnetic moment of the cloud $\mu =\int
\delta\mu_z(r)d^2r$  is convergent at large
distances, and it means that the above assumption is valid.
This implies that {\it all} higher order in $J_{\perp}$ infrared divergent
diagrams must cancel out. This is a highly unusual situation and it would be
interesting to check the cancellation by a direct calculation.

\section{conclusions}
We have analyzed excitation spectrum  of a finite
antiferromagnetic cluster with magnetic impurity, and considered a crossover
between quantum and classical Curie law for the impurity magnetic
susceptibility. We have also derived a logarithmic correction to the impurity
magnetic susceptibility. Dependent on the parameters this can be logarithm of
system size, temperature, or frequency of the external magnetic field.
Using the results for the logarithmic correction we have derived the Ward identity
for the  impurity-spin-wave vertex.

\section{acknowledgments}
I am very grateful to A. W. Sandvik and G. Khaliullin for important
 discussions.  I am also very grateful to  K. H. H\"oglund, A. W. Sandvik,
and S. Sachdev for communicating me their results prior publication.


\begin{references}
\bibitem{Man} E. Manousakis, Rev. Mod. Phys. {\bf 63}, 1 (1991).
\bibitem{Dag} E. Dagotto, Rev. Mod. Phys. {\bf 66}, 763 (1994).
\bibitem{Cheong}S. -W. Cheong {\it et al}., Phys. Rev. B {\bf 44}, 9739 (1991);
S. T. Ting {\it et al., ibid.} {\bf 46}, 11772 (1992);
M. Corti  {\it et al., ibid.} {\bf 52}, 4226 (1995);
P. Carretta, A. Rigamonti, and R. Sala, {\it ibid.}
 {\bf 55}, 3734 (1997).
\bibitem{Vajk} O. P. Vajk, P. K. Mang, M. Greven, P. M. Gehring, and
  J. W. Lynn, Science {\bf 295}, 1691 (2002).
\bibitem{Nag} N. Nagaosa, Y. Hatsugai, and M. Imada, J. Phys. Soc. Jpn. 
{\bf 58}, 978 (1989);
N. Nagaosa and T. -K. Ng, Phys. Rev. B {\bf 51}, 15588 (1995).
\bibitem{Bulut} N. Bulut, D.Hone, D. J. Scalapino, and E. Y. Loh,
  Phys. Rev. Lett. {\bf 62}, 2192 (1989).
\bibitem{Sand1} A. W. Sandvik, E. Dagotto, and D. J. Scalapino, Phys. Rev. B
  {\bf 56}, 11701 (1997). 
\bibitem{Sachd} S. Sachdev, C. Buragohain, and M. Vojta, Science {\bf 286},
  2479 (1999);
M. Vojta, C. Buragohain, and S. Sachdev, Phys. Rev. B {\bf 61}, 15152 (2000).
\bibitem{Sand2} K. H. H\"oglund and A. W. Sandvik, cond-mat/0302273.
\bibitem{Igar} J. Igarashi, K. Murayama, and P. Fulde, Phys. Rev. B {\bf 52},
15966 (1995); K. Murayama and J. Igarashi, J. Phys. Soc. Jpn. 
{\bf 66}, 1157 (1997).
\bibitem{Kotov} V. N. Kotov, J. Oitmaa, and O. P. Sushkov,  Phys. Rev. B {\bf 58},
8495 (1998).
\bibitem{Chern} A. L. Chernyshev, Y. C, Chen, and A. H. Castro Neto,
  Phys. Rev. B {\bf 65}, 104407 (2002).
\bibitem{Sus} O. P. Sushkov, Phys. Rev. B {\bf 62}, 12135 (2000).
\bibitem{Troy} M. Troyer, Prog. Theor. Phys. Suppl. {\bf 145}, 326 (2002).
\bibitem{Neub}H. Neuberger and T. Ziman, Phys. Rev. B {\bf 39}, 2608 (1989).
\bibitem{Runge} K. J. Runge,  Phys. Rev. B {\bf 45}, 12292 (1992).
\bibitem{Has} P. Hasenfratz and F. Niedermayer, Z. Phys. B {\bf 92}, 91 (1993).
\bibitem{Singh} R. R. P. Singh, Phys. Rev. B {\bf 39}, 9760 (1989).
\bibitem{Zheng} Zheng Weihong, J, Oitmaa, and C . J. Hamer, Phys. Rev B
{\bf 43}, 8321 (1991).
\bibitem{Land} L. D. Landau and E. M. Lifshitz. Quantum Mechanics:
  non-relativistic theory. Oxford, New York, Pergamon Press, 1977.
\bibitem{Sachd1} S. Sachdev, M. Vojta, cond-mat/0303001.



\end{references}
\end{document}